\documentclass[prl,aps,twocolumn, showpacs,superscriptaddress,groupedaddress]{revtex4}

\usepackage{graphicx}
\usepackage{amsmath}
\usepackage{bm}
\usepackage[latin1]{inputenc}
\usepackage{amstext}
\usepackage{amsxtra}
\usepackage{times}
\usepackage{xspace}
\usepackage{xcolor}
\usepackage{setspace}
\usepackage{float}
\usepackage{varioref}
\usepackage{hyperref}
\newcommand{\bs}{\boldsymbol}
\newcommand{\beq}{\begin{equation}}
\newcommand{\eeq}{\end{equation}}
\newcommand{\bpm}{\begin{pmatrix}}
\newcommand{\epm}{\end{pmatrix}}
\newcommand{\bea}{\begin{eqnarray}}
\newcommand{\eea}{\end{eqnarray}}

\newcommand{\kb}{k_{\rm B}}
\newcommand{\lt}{\lambda_{\rm T}}

\newcommand{\gt}{\ensuremath{\tilde{g}}\xspace}

\newcommand{\Rb}{$^{87}$Rb \xspace}
\newcommand{\psd}{\ensuremath{\mathcal{D}}\xspace}
\newcommand{\mcp}{\ensuremath{\mathcal{P}}\xspace}

\newcommand{\hfmf}{Hartree--Fock mean-field\xspace}

\newcommand{\Is}{I_{\rm sat}}

\newcommand{\eos}{{\sc E}o{\sc S}\xspace}

\usepackage{times}

\begin{document}
\title{Fit-free determination of scale invariant equations of state:\\ application to the 2D Bose gas across the Berezinksii--Kosterlitz--Thouless transition}
\date{\today}

\author{R\'emi Desbuquois}
\affiliation{Institute for Quantum Electronics, ETH Zurich, 8093 Zurich, Switzerland}

\author{Tarik Yefsah}
\affiliation{MIT-Harvard Center for Ultracold Atoms, Research Laboratory of Electronics, Department of Physics, Massachusetts Institute of Technology, Cambridge, Massachusetts 02139, USA}

\author{Lauriane Chomaz}
\affiliation{Laboratoire Kastler Brossel, CNRS, UPMC, ENS, 24 rue Lhomond, F-75005 Paris, France}
\affiliation{Coll\`ege de France, 11 Place Marcelin Berthelot, 75231 Paris Cedex 05, France}

\author{Christof Weitenberg}
\affiliation{Institut f\"ur Laserphysik, Universit\"at Hamburg, Luruper Chaussee 149, D-22761 Hamburg, Germany}

\author{Laura Corman}
\affiliation{Laboratoire Kastler Brossel, CNRS, UPMC, ENS, 24 rue Lhomond, F-75005 Paris, France}

\author{Sylvain Nascimb\`ene}
\affiliation{Laboratoire Kastler Brossel, CNRS, UPMC, ENS, 24 rue Lhomond, F-75005 Paris, France}

\author{Jean Dalibard}
\affiliation{Laboratoire Kastler Brossel, CNRS, UPMC, ENS, 24 rue Lhomond, F-75005 Paris, France}
\affiliation{Coll\`ege de France, 11 Place Marcelin Berthelot, 75231 Paris Cedex 05, France}

\begin{abstract}
We present a general ``fit-free" method for measuring the equation of state (\eos) of a scale-invariant gas. This method, which is inspired from the procedure introduced by Ku et al. [Science 335, 563 (2012)] for the unitary three-dimensional Fermi gas, provides a general formalism which can be readily applied to any quantum gas in a known trapping potential, in the frame of the local density approximation. We implement this method on a weakly-interacting two-dimensional Bose gas in the vicinity of the Berezinskii-Kosterlitz-Thouless transition, and determine its \eos with unprecedented accuracy in the critical region. Our measurements provide an important experimental benchmark for classical field approaches which are believed to accurately describe quantum systems in the weakly interacting but non-perturbative regime.
\end{abstract}
\maketitle

Homogeneous matter at thermal equilibrium is described by an equation of state (\eos), \emph{i.e.,} a functional relation between thermodynamic  variables of the system. While the \eos is analytically known for ideal gases, one must resort to approximations or numerical calculations to determine the \eos of interacting fluids, which must then be confronted to experiments. Thanks to a precise control of temperature, confining potential and interaction strength, cold atomic gases constitute a system of choice for the experimental determination of quantum matter \eos \cite{Ho:2009}. While performed on atomic systems, such measurements often provide crucial insight on generic physical problems, well beyond the atomic physics perspective. Prominent examples are the recent measurements of the \eos of atomic Fermi gases \cite{Nascimbene:2010,Navon:2010,Ku:2012,VanHoucke:2012}, which provided a precious quantitative support for our understanding of strongly interacting fermions at low temperature. Another important paradigm accessible to atomic gases is found in two-dimensional quantum systems, where the low temperature state is established via a defect-driven transition. This generic phenomenon of two-dimensional systems is described by the celebrated Berezinskii-Kosterlitz-Thouless (BKT) theory, with a scope that ranges from superconductivity to quantum Hall bilayer physics to high energy physics.

In this context, the weakly interacting two-dimensional Bose gas is of particular interest as it supports the fundamental principles of the BKT theory, while allowing for a simplified theoretical description. Indeed, for small enough inter-particle interactions, the dynamics of the two-dimensional Bose gas is well captured by a classical field model~\cite{Prokofev:2001,Prokofev:2002}, which is itself described by a dimensionless coupling constant and exhibits scale-invariance \footnote{Beyond the classical field level the scale invariance is broken~\cite{Vogt:2012}, but the deviation can be neglected for small enough interaction strengths~\cite{Olshanii:2010,Langmack:2012}}. In general, scale invariance occurs in any fluid where no explicit energy/length scale is associated to the (binary) interaction potential. For the weakly interacting two-dimensional case, the length scale introduced by the 3D scattering length is normalized by the extension of the system in the third dimension, and this dimensionless ratio characterizes the effective 2D interaction strength. Scale invariance is also found in the unitary Fermi gas, where the scattering length describing $s-$wave interactions diverges (for a review, see \cite{Giorgini:2008}). This property considerably simplifies the \eos structure, as general dimensionless quantities such as the phase space density \psd which usually depend separately on the chemical potential $\mu$ and the temperature $T$ can only be expressed as the ratio $\mu /\kb\,T$ owing to the absence of other energy scales \cite{Hung:2011,Yefsah:2011}.

The usual method for determining the \eos of a cold atomic gas starts with the measurement of the density distribution $n(\bs r)$ in a smoothly varying, confining potential $V(\bs r)$. The gas is assumed to be in thermal equilibrium, and its chemical potential at the trap center is $\mu$. Using  the Local Density Approximation (LDA), the measured ${\cal D}(\bs r)$ is linked to that of a uniform fluid with the same interaction strength, the same temperature, and with chemical potential $\mu(\bs r)=\mu - V(\bs r)$ \cite{Ho:2009}. The values of $T$ and $\mu$ for a given realization of the gas are obtained by comparing the low-density wings of $n(\bs r)$ with the known theoretical result for a dilute fluid. A major drawback of this method is the fact that any systematic error in the determination of the density, \emph{e.g.} due to imperfect calibration of the probing system, will lead to inaccurate values of $\mu$ and $T$, and thus affect the measurement of the \eos ${\cal D}(\mu,T)$. Recently, an alternative fit-free method that does not suffer from this limitation has been put forward in \cite{Ku:2012} for the measurement of the EoS of the scale-invariant 3D Fermi gas. It is based on the use of two specific thermodynamic variables, pressure and compressibility; in addition, absolute energy scales $T$ and $\mu$ were replaced by a single relative scale ${\rm d}\mu$, which was itself determined by the LDA through ${\rm d}\mu=-{\rm d}V$.

The purpose of the present Letter is two-fold. First, we describe a method that generalizes the procedure introduced in \cite{Ku:2012}, which does not rely on specific thermodynamic variables but rather provides a generic formalism that can readily be applied to other quantum systems. Second, we implement this method on a two-dimensional  (Bose) fluid, for which all information about the spatial density $n(\bs r)$ is directly accessible from an image of the cloud. This is to be contrasted with the 3D case studied in \cite{Ku:2012}, where one has to resort to an Abel transform to reconstruct the spatial density profile, before implementing the fit-free determination of the \eos. The \eos reconstructed here reaches a precision which makes it suitable for a quantitative comparison to the classical-field Monte Carlo calculation \cite{Prokofev:2002} in the critical region, which could not be conclusive from previous measurements~\cite{Hung:2011,Yefsah:2011}. Our measurement, with a relative statistical error smaller than 1\% on the detectivity, is found in excellent agreement (better than 5\%) with the prediction obtained from \cite{Prokofev:2002} at the critical point and deeper in the superfluid regime. In the normal regime close to the transition point, we observe a deviation on the order of 15\%, which might be due to beyond classical-field effects.

We start our analysis by considering an atomic gas in thermodynamic equilibrium confined in a known potential $V(\bs r)$. The only hypothesis for the method is the LDA, which entails that $n(\bs r)$ depends on position only through the local value of the trapping potential: $n(\bs r)=n[V(\bs r)]$. Although this method is applicable to any dimension, we focus here on the particular case of the two-dimensional gas for the sake of clarity. Let us introduce the energy $E[V(\bs r))]$ with $\bs r=(x,y)$, defined by \footnote{In dimension $d$, this equation takes the general form $E=\hbar^2\,n^{2/d}/m$} 
\begin{equation}
E=\frac{\hbar^2}{m} n,
\label{eq:En}
\end{equation}
which we want to combine with other relevant energies in order to form useful dimensionless variables. Though no absolute energy scales are readily available, a relative energy scale is provided by the variation of the trapping potential ${\rm d}V$. Furthermore, quantities formed in this manner are directly connected by the LDA to the properties of the uniform gas using the relation ${\rm d}\mu=-{\rm d}V$. Thus, we define the dimensionless quantities
\begin{equation}
X_\nu\equiv E^{\nu-1}\frac{\partial^\nu E}{\partial \mu^\nu}=(-1)^\nu\,E^{\nu-1}\frac{{\rm d}^\nu E}{{\rm d} V^\nu}\,,
\label{eq:Xnu}
\end{equation}
where $\nu$ is an integer. By convention, $X_0=1$ and a negative $\nu$ will instead correspond to $|\nu|$ successive integrations of $E$ with respect to $V$, with for example
\begin{equation}
X_{-1}=\frac{1}{E^2}\int^{\infty}_V E(V')\; {\rm d}V'\,.
\label{eq:Xm1}
\end{equation}
From a given image of the gas $n(\bs r)$, one can thus construct all functions $X_\nu(V)$. In the case of a scale invariant system, the knowledge of a single thermodynamic variable $X_\nu$ is sufficient to determine the state of the fluid, hence the values of all other variables $X_{\nu'}$. In other words,  all individual measurements must collapse on a single line in each plane $\{X_\nu,\,X_{\nu'}\}$, irrespective of their temperature and  chemical potential. Such a line is a valid EoS of the fluid under consideration.

Once the $X_\nu$ are known, all other thermodynamic quantities can be determined, up to an integration constant. In particular, one can derive the phase-space density \psd and the ratio $\alpha=\mu/\kb T$. Let us suppose that a point $(X_\nu^{(0)},\,X_{\nu'}^{(0)})$ can be identified in a known portion the \eos, and that it corresponds to the values $\alpha_0$ and $\psd_0$. The link between the set $\{X_\nu\}$ and $(\alpha,\psd)$ is provided by
\begin{align}
\label{eq:genPsd}
\psd(X_\nu^{(1)})=\psd_0\exp\left(\int_{X_\nu^{(0)}}^{X_\nu^{(1)}}\frac{X_1}{(\nu-1)X_1X_\nu+X_{\nu+1}}{\rm d}X_\nu\right),\\
\label{eq:genAlpha}
\alpha(X_\nu^{(1)})=\alpha_0+\frac{1}{2\pi}\int_{X_\nu^{(0)}}^{X_\nu^{(1)}}\frac{\psd(X_\nu)}{(\nu-1)X_1X_\nu+X_{\nu+1}}{\rm d}X_\nu\,.
\end{align}
The determination of $(\alpha,\psd)$ thus requires the knowledge of a triplet $\{X_1,X_\nu,X_{\nu+1}\}$. This requirement can be weakened by choosing $\nu=1$ or $\nu=-1$, in which case only the pairs $\{X_1,X_{-1}\}$ or $\{X_1,X_{2}\}$ are needed, and  $(\alpha,\psd)$ are given by 
\begin{align}
\label{eq:Psd}
\psd(X_{-1}^{(1)})=\psd_0\exp\left(\int_{X_{-1}^{(0)}}^{X_{-1}^{(1)}}\frac{X_1}{1-2X_1X_{-1}}{\rm d}X_{-1}\right)\,,\\
\label{eq:Alpha}
\alpha(X_{-1}^{(1)})=\alpha_0+\frac{1}{2\pi}\int_{X_{-1}^{(0)}}^{X_{-1}^{(1)}}\frac{\psd(X_{-1})}{1-2X_1X_{-1}}{\rm d}X_{-1}\,,
\end{align}
or 
\begin{align}
\psd(X_1^{(1)})=\psd_0\exp\left(\int_{X_1^{(0)}}^{X_1^{(1)}}\frac{X_1}{X_{2}}{\rm d}X_1\right) \,,\\
\alpha(X_1^{(1)})=\alpha_0+\frac{1}{2\pi}\int_{X_1^{(0)}}^{X_1^{(1)}}\frac{\psd(X_1)}{X_{2}}{\rm d}X_1.
\end{align}

\begin{figure}[t]
\centering
\includegraphics[width=80mm]{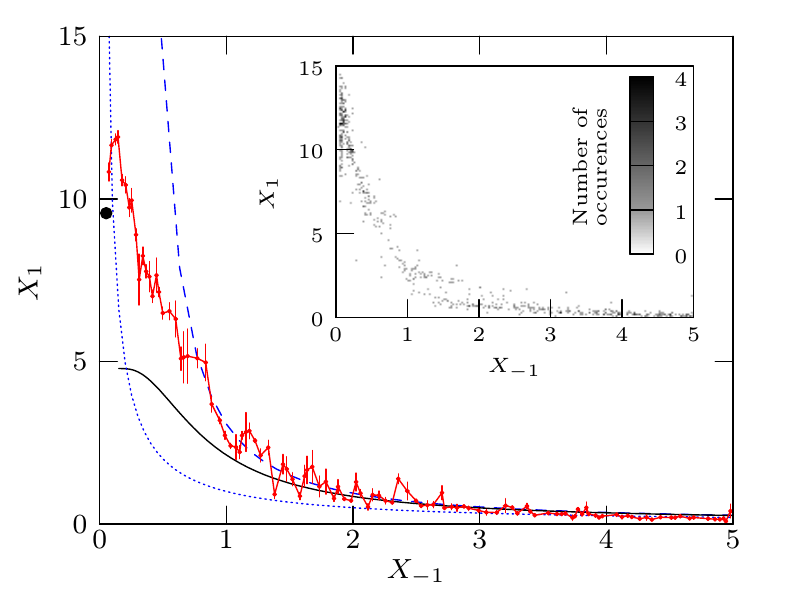}
\caption{(Color online) Determination of the \eos with variables $X_{-1}$ and $X_1$, along with known limits for the \eos. The simple cases of the ideal Bose gas (Boltzmann gas) is shown in blue dashed (dotted) line. The known limits of the \eos of the weakly interacting 2D Bose gas are indicated by a black point for the Thomas--Fermi limit and by a black full line for the Hartree--Fock mean field theory.  The red line results from the averaging over all the separate intensity profiles, with the error bars corresponding to the standard error introduced by the averaging procedure. The data shown here contains $\sim$ 100 different values of $X_{-1}$. Inset:  distribution of measured values of $X_{-1}$ and $X_1$. The gray level indicates the number of individual data points falling in each pixel.}
\label{fig:X1Xm1}
\end{figure}

We illustrate this general procedure with a few examples. For the simple case of a Maxwell-Boltzmann gas, the \eos in terms of the $X_\nu$'s can be obtained analytically and one gets for example $X_1 X_{-1}=1$ and $X_2=X_1^2$. For an interacting 2D gas, the \eos  is not known analytically; however for the bosonic case, it can be approximated in two limiting cases. For $\mu<0$, the gas is only weakly degenerate and the mean-field energy of an atom in the gas can be written $2\tilde g (\hbar^2 n/m)$, where the dimension-less coefficient $\tilde g$ (assumed here $\ll 1$) characterizes the strength of the interaction. The thermodynamics is then well described by the prediction of the  Hartree--Fock theory \cite{Hadzibabic:2011}
\begin{equation}
\psd=-\ln\left( 1 - e^{\alpha-\gt \psd/\pi}\right),
\end{equation}
from which we extracted numerically the values of $X_{-1}$ and $X_1$ and plotted the corresponding \eos  in Fig.~\ref{fig:X1Xm1}. 
In the opposite case of a strongly degenerate gas (chemical potential positive and larger than $\kb T$), the gas is described by the Thomas--Fermi equation $\psd=2\pi\,\alpha/\gt$. All $X_\nu$ are then constant, with $X_{-1}=\gt/2$, $X_1=1/\gt$ and $X_2=0$.

We now turn to the practical implementation of this method for processing data obtained with a quasi-2D rubidium gas. Our experimental preparation follows the lines detailed in \cite{Rath:2010,Yefsah:2011}. We start with a 3D gas of \Rb atoms, confined in their $F=m_F=2$ state in a magnetic trap. To create a 2D system, we shine an off-resonant blue-detuned laser beam on the atoms, with an intensity node in the plane $z=0$. The resulting potential provides a strong confinement perpendicular to this plane, with oscillation frequency $\omega_z/2\pi=1.9\,(2)$ kHz, which decreases at most by 5\% over typical distribution radii. This corresponds to the interaction strength $\gt=\sqrt{8\pi}\,a/l_z\approx 0.1$, where $a$ is the 3D scattering length and $l_z=\sqrt{\hbar/m\omega_z}$ \cite{Petrov:2000a}. The energy $\hbar\omega_z$ is comparable to the thermal energy $\kb T$, which ensures that most of the atoms occupy the ground state of the potential along $z$ (see \cite{Yefsah:2011} and below). The magnetic trap provides a harmonic confinement in the $xy$ plane, with mean oscillation frequency $\omega_r/2\pi=20.6\,(1)$ Hz.

After letting the cloud equilibrate for 3\,s in this combined magnetic and dipole trap, we measure the density distribution $n(\bs {r})$ by performing \emph{in-situ} absorption imaging with a probe beam perpendicular to the plane of the atoms. In order to overcome the multiple scattering effects stemming from the high spatial density of our sample \cite{Rath:2010,Chomaz:2012}, the intensity of the probe beam $I$ is chosen to be much larger than the saturation intensity $\Is$ of the $D_2$ resonance line of Rb (typically, $I/\Is\approx 40$). In this regime, the response of individual atoms is independent of their environment \cite{Reinaudi:2007}. Owing to the photonic shot-noise, this method inherently suffers from a low signal-to-noise ratio. For this reason, we complement it by the conventional low-intensity absorption imaging ($I/\Is\approx 0.7$). In this regime, the noise level is much lower, though high atomic densities cannot be measured. Since both imaging processes destroy the atomic distribution, each sample is prepared twice and imaged successively in the low and high intensity regime. For the data analyzed below, we used $\approx 80$ images, with temperatures ranging from 30\,nK to 150\,nK, and atoms numbers from 25\,000 to 120\,000. 

\begin{figure}[tr]
\centering
\includegraphics[width=80mm]{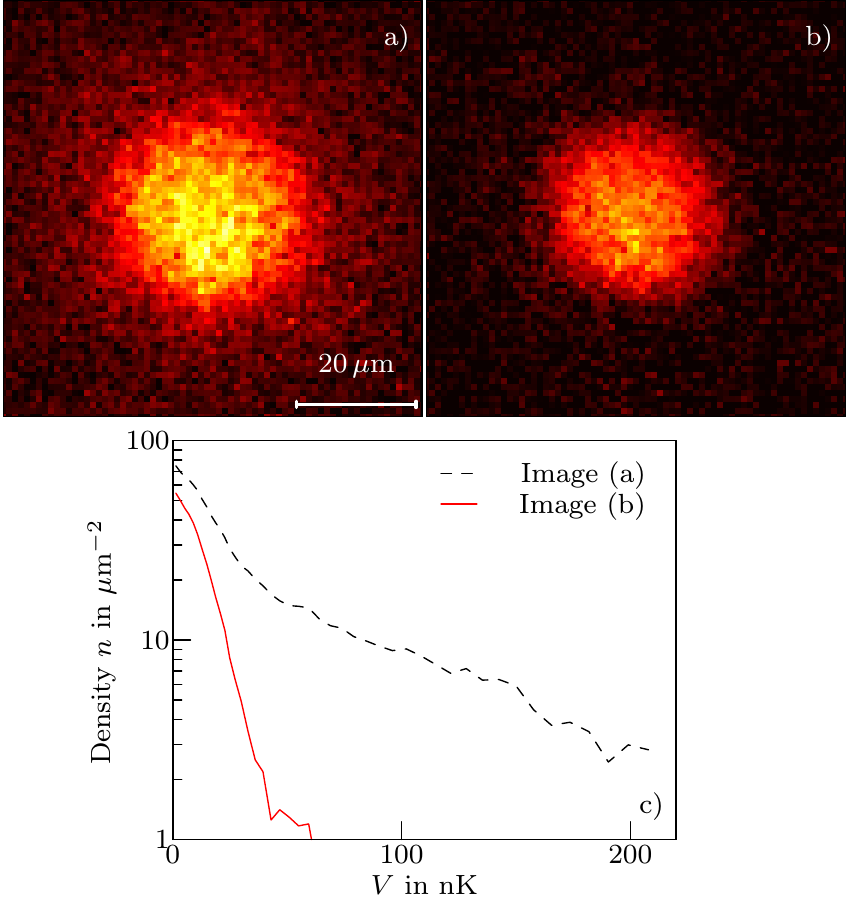}
\caption{(Color online) (a) and (b) Density distributions of 2D atomic samples of $^{87}$Rb corresponding to a partially degenerate (a) and a strongly degenerate cloud (b). (c) Corresponding function $n[V(\bs r)] $resulting from azimuthal averaging. The distributions are obtained with high intensity imaging.}
\label{fig:HotVsCold}
\end{figure}

In Fig.~\ref{fig:HotVsCold} we show typical density distributions of 2D atomic clouds, together with the corresponding function $n[V(\bs r)]$. The cloud (a) exhibits a significant thermal fraction, contrarily to cloud (b), which is essentially in the Thomas-Fermi regime. The latter illustrates the power of this fit-free method, since it can be incorporated as such in our determination of the \eos. On the opposite, it would be discarded in a conventional approach, owing to the impossibility to assign it a temperature.

Though both choices of variables $(X_{-1},X_1)$ and $(X_1,X_2)$ are in principle possible, the latter requires the experimental evaluation of a second-order derivative, which often suffers from a poor signal-to-noise ratio. By contrast the choice $(X_{-1},X_1)$, also adopted in \cite{Ku:2012} when writing the \eos in terms of pressure and compressibility, appears particularly robust \footnote{In our case, the compressibility is given by $\kappa=X_1/(n^2\,h^2/m)$ and the pressure by $P=X_{-1}\,n^2\,h^2/m$}. For each image, we perform an azimuthal average and compute a set of $\approx 70$ data points $(X_{-1},\,X_1$), where the low (high) values of $X_{-1}$ correspond to the high (low) density regions of the image. 

In a first step, we combine all sets obtained from images acquired at various temperatures and various atom numbers to test the scale invariance. As explained above, each individual measurement should sit on the same universal curve in the $(X_{-1},X_1)$ plane, provided the interaction strength $\tilde g$ is constant. We show in the inset of Fig.~\ref{fig:X1Xm1} the repartition of data points in the $(X_{-1},X_1)$ plane, which fall as expected around a single curve. 

After having checked the (approximate) scale invariance of our quasi-2D Bose gases, we average all data sets  $(X_{-1},X_1)$ obtained from the ensemble of images and  arrive at the determination of the \eos  shown in Fig.~\ref{fig:X1Xm1} \footnote{We extract data points with $X_{-1}<2.5$ ($X_{-1}>2.5$)  from high (low) intensity imaging.}. In order to re-express this \eos in terms of the more traditional variables $\alpha$ and \psd, we now need to apply the transformations of Eqs. (\ref{eq:Psd}) and (\ref{eq:Alpha}). However, this transformation must be adapted to account for possible imperfections in the calibration of the detectivity of our imaging setup. Indeed, as in most cold atoms experiments, we only measure the density up to a global multiplicative factor $\beta$ \footnote{This factor is taken into account in Eq. \ref{eq:En} by $E=\beta( \hbar^2/m)n$. The definition of $X_\nu$ is unchanged.}, which is defined as the ratio between the unknown actual absorption cross-section and the ideal one expected for monochromatic probe light in the absence of stray magnetic fields. Taking this calibration factor into account amounts to replacing Eqs. (\ref{eq:genPsd}) and (\ref{eq:genAlpha}) by  
\begin{align}
\psd(X_\nu^{(1)})=\psd_0\exp\left(\int_{X_\nu^{(0)}/\beta^\nu}^{X_\nu^{(1)}/\beta^\nu}\frac{X_1}{(\nu-1)X_1X_\nu+X_{\nu+1}}{\rm d}X_\nu\right)\\
\alpha(X_\nu^{(1)})=\alpha_0+\frac{\beta}{2\pi}\int_{X_\nu^{(0)}/\beta^\nu}^{X_\nu^{(1)}/\beta^\nu}\frac{\psd(X_\nu)}{(\nu-1)X_1X_\nu+X_{\nu+1}}{\rm d}X_\nu
\end{align}
 where the bounds of the integrals now depend on $\beta$ and where $X_\nu^{(0)}/\beta^\nu$ corresponds to the reference values $\alpha_0$ and $\psd_0$. The value of $\beta$ is \emph{a priori} unknown; however, it can be determined by performing a least-square fit of the prediction of the \hfmf theory to the measured \eos over a proper region of phase-space densities. In other words, the fact that the value of $\beta$ is unknown can be handled in this method, provided one has a good theoretical knowledge of the \eos over a segment in parameter space, rather than at a single point $(\alpha_0,\psd_0)$. Here, we consider the \eos to be well approximated by the \hfmf theory for $\alpha<0$.

\begin{figure}[t]
\includegraphics[width=80mm]{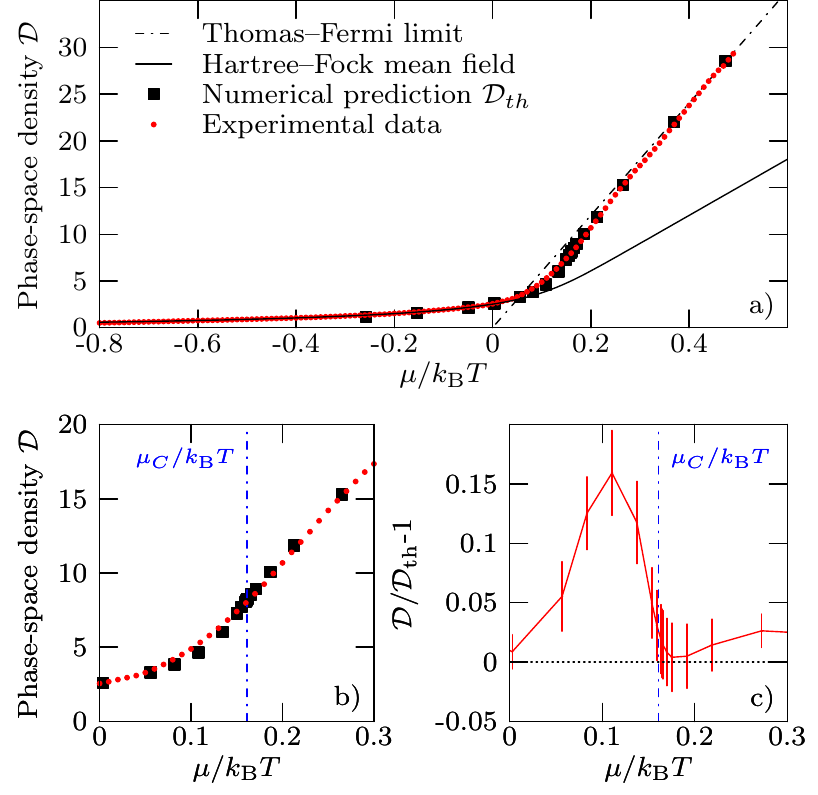}
\caption{(Color online) (a) Equation of state of the 2D Bose gas, determined with Eqs. \ref{eq:Psd} and \ref{eq:Alpha} (red points), with a detailed view of the critical region around the BKT transition (blue dash-dotted line) in (b). Statistical error bars are too small to be shown on these plots. We show for comparison the classical field Monte Carlo prediction $\psd_{\rm th}$ \cite{Prokofev:2002} in black squares, the Thomas--Fermi limit in black dash-dotted line and the \hfmf theory in black full line. We provide a quantitative estimate of the difference between measurement and prediction in (c). There, we plot $\psd/\psd_{\rm th}-1$, where zero indicates perfect agreement. The error bars result from a boostrap analysis of the experimental data.} 
\label{fig:eos}
\end{figure}

We choose the central point $(\alpha_0,\psd_0)$ of the fitting segment  at $X_{-1}^0=3$, which corresponds to a phase-space density $\psd_0=1.45$ and $\alpha_0=-0.22$, well within the \hfmf regime. This choice is a good compromise between higher values of $X_{-1}$, which have a significantly higher noise level and lower values of $X_{-1}$, for which the \eos is not known. The fit of the \hfmf theory yields a detectivity $\beta=0.456\,(1)$ \footnote{The error bar on this value is calculated through a bootstrap analysis. The result is consistent with our (less precise) previous estimate for the same setup, $\beta = 0.40 (2)$, which was based on a different method \cite{Yefsah:2011}.}. The \eos in terms of the variables $(\alpha,\psd)$ -- obtained after a small correction due to excited states of the $z$ motion (see below) --  is shown in Fig.~\ref{fig:eos}a, along with the numerical prediction $\psd_{\rm th}$ \cite{Prokofev:2002}. Note that the reconstructed \eos is remarkably smooth and doesn't display any particular feature at the transition point. This observation is also made on the \eos for pressure, entropy and heat capacity shown in the Supplementary Material. This fact illustrates the ``infinite order" nature of the BKT transition which is not associated to any singularity of thermodynamic quantities\footnote{A singularity however appears in the superfluid density, which has to be distinguished from the total density, as the former enters into account only in the response of the system to an externally driven drag or rotation.}, as opposed to phase transitions driven by the breaking of a continuous symmetry, such as the second order lambda transition observed at MIT \cite{Ku:2012}. To compare quantitatively the reconstructed \eos with the numerical prediction, we plot the quantity $\psd/\psd_{\rm th}-1$ in Fig.~\ref{fig:eos}, and find it to lie consistently below 15 \%, and even below 5 \% around the phase transition, which occurs at $\mu_C/\kb T\approx 0.17$ \cite{Prokofev:2002}. The deviation observed in the fluctuation region below the critical chemical potential might signal deviations to the classical field picture which is expected to be accurate for $\gt\ll\,1$ \cite{Prokofev:2001,Prokofev:2002}. Theoretically this deviation could be addressed using Quantum Monte Carlo methods \cite{Holzmann:2008,Rancon:2012}.

We have emphasized that the above method does not require any fit of individual images, nor the knowledge of their temperature $T$ or chemical potential $\mu$. Furthermore, once the \eos has been determined, it is possible to fit the full \eos to the individual images and to infer their $(T,\mu)$. We find from this \emph{a posteriori} fit that the temperatures of our gases range from 30 nK to 150 nK, which is of the same order as the harmonic oscillator level spacing $\hbar\omega_z/\kb \sim 100$~nK. Consequently, our warmest clouds do not  strictly lie in the 2D regime; some atoms occupy excited states of the $z$ motion, which breaks the scale invariance of the \eos. Fortunately, as a consequence of Bose statistics, this fraction of atoms remains small even when $\kb T \sim \hbar \omega_z$. Their contribution to the total density can then be self-consistently evaluated and removed, in a similar manner to  \cite{Tung:2010,Yefsah:2011}. The remaining ground-state density can then be used to refine the determination of the \eos. In the data presented in Fig.~\ref{fig:eos}, this subtraction of the contribution of the excited states of the $z$ motion has already been performed. The exact procedure used for the excited state subtraction is presented in detail in the Supplemental Material, together with the initial \eos obtained before this subtraction (the deviations with respect to the final one shown in Fig.~\ref{fig:eos} are minute).

In conclusion, we have presented a method to determine the \eos of a scale-invariant fluid. It is based on the identification of a set of dimensionless variables $X_\nu$ directly related to the measured density of the fluid. This method does not rely on thermometry of individual images, nor on the precise calibration of the detectivity, and leads to a strong reduction of the noise level in the measurement. We have applied it to the case of a weakly interacting Bose gas and obtained its \eos with a  precision of few percents, in excellent agreement with a theoretical prediction obtained from a classical Monte Carlo simulation. Using the response of the gas to a gauge field, originating for example from a rotation, this method could be extended to access the superfluid faction of the gas along the lines proposed in \cite{Ho:2009}. In principle, this method is not limited to scale invariant systems, and could be extended to any situation described by two independent dimensionless parameters, such as the zero temperature limit of the Fermi gas, either for a spin-balanced gas with varying interactions \cite{Navon:2010}, or for a unitary spin-imbalanced Fermi gas \cite{Shin:2008}.

\begin{acknowledgments}
We thank J\'er\^ome Beugnon, Jason Ho, Boris Svistunov and Martin Zwierlein for useful discussions and comments. This work is supported by IFRAF, ANR (ANR-12-BLAN-AGAFON), ERC (Synergy UQUAM). L. Chomaz and L. Corman acknowledge the support from DGA, and C. Weitenberg acknowledges the support from the EU (PIEF-GA-2011-299731).
 \end{acknowledgments}

\bibliography{References-eos}

\newpage
\renewcommand\thefigure{S\arabic{figure}}
\setcounter{figure}{0}   
\centerline{\Large \textbf{Supplemental material}}

\vskip 5 mm

\paragraph{Removing the contribution for thermally populated transverse states} The method presented in the main text assumes the \eos to be strictly scale invariant. However, when the temperature $T$ of an individual realization is on the order of the harmonic oscillator level spacing $\hbar\omega_z/\kb \sim 100$~nK, a fraction of the atoms occupy transverse excited states of the confinement potential. Their contribution can be evaluated and self-consistently removed in the following manner.

\begin{enumerate}
	\item We calculate the \eos in variables $(X_{-1},X_1)$ with the method described in the main text. This process yields a preliminary determination of the \eos, as well as a value for the detectivity $\beta_{(0)}=0.481(4)$. Both these measurements contain a systematic error introduced by the thermally populated transverse levels. 
	\item We measure the temperature and the chemical potential of the atomic distribution of each individual image by fixing the detectivity $\beta$ to the value determined above, and by fitting with the \eos measured previously. Since the \eos contains excited levels in its determination, we expect that the measured temperature will also be affected. Note that the population of the excited states is only 10\% of the total population at most: we therefore expect a similarly small shift of the temperature and chemical potential. 
	\item We self-consistently evaluate the contribution of the excited levels to the total density, assuming the atoms in the excited states of the $z$ motion to be in the HFMF regime \cite{Tung:2010,Yefsah:2011} and subtract them to obtain an estimate of the population of the ground state. We find that the phase-space density associated with the excited states is at most 0.5, which justifies to describe these states with the \hfmf approximation.	
	\item With the new estimate of the population of the ground state for each image, we generate a new \eos with the	method outlined in the main text. We also determine\begin{figure}[H]
\includegraphics[width=8cm]{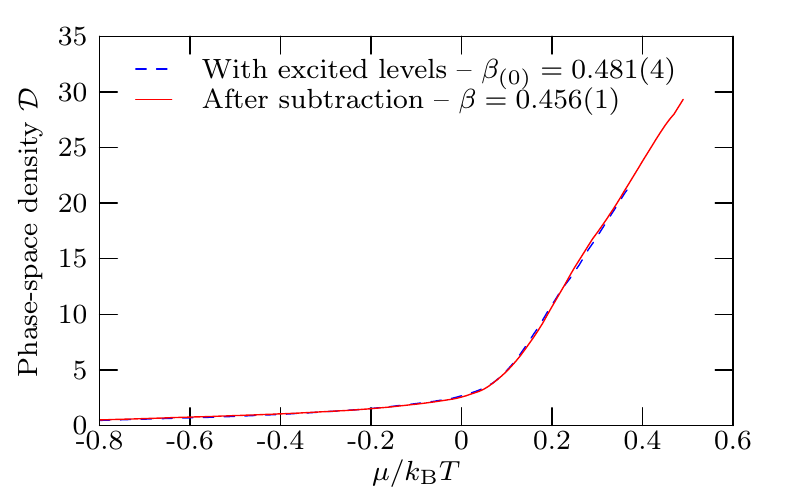}
\caption{(Color online) \eos of the 2D Bose gas, determined with Eqs. \ref{eq:Psd} and \ref{eq:Alpha}. The measurement without (with) subtraction of the population of the excited levels is shown in blue (red) line.}%
\label{fig:eos_supp}%
\end{figure} a more accurate value of the detectivity, and find $\beta=$$0.456(1)$.
	
  \item In principle, this process can be iterated to obtain an even more accurate determination of the temperature for each image. However, this does not lead to an improved measurement.
\end{enumerate}
As shown in Fig.~\ref{fig:eos_supp}, the subtraction of the excited states does not significantly affect the \eos, though the detectivity is significantly modified.
\begin{figure}[b]
\includegraphics[width=8cm]{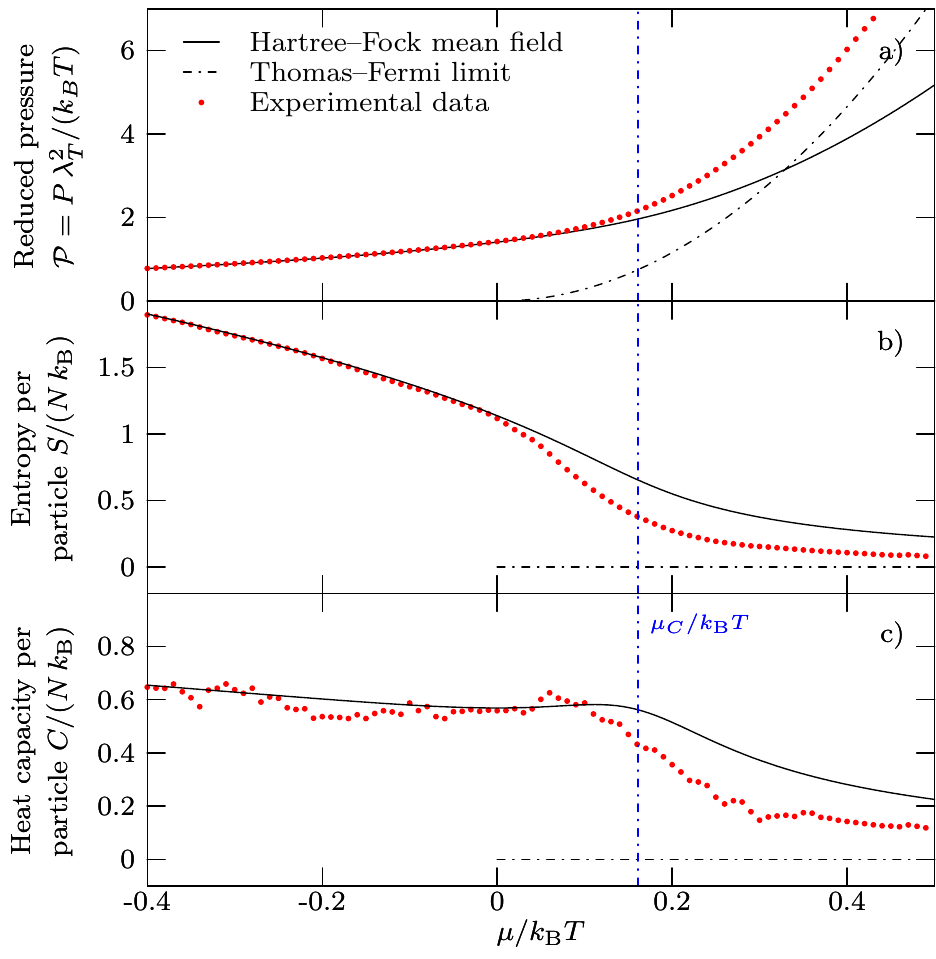}
\caption{(Color online) Equations of state of the 2D Bose gas, derived from the \hfmf theory (black line) and reconstructed following our method (red points), expressed with (a) the pressure $P$, (b) the entropy per particle $S/N$ and (c) the heat capacity per particle $C/N$.}%
\label{fig:eos2_supp}%
\end{figure}
\paragraph{EoS in other variables} Owing to the scale invariance of the two dimensional Bose gas, the \eos with any pair of variables can in principle be deduced from the one in the main text. We show in Fig.~\ref{fig:eos2_supp} three measurements of such \eos. For example, the pressure is deduced from the density by $n=(\partial P/\partial\mu)_T$ and is expressed in dimensionless units:
\begin{equation}
\mcp=\frac{P\,\lt^2}{\kb}=\frac{1}{2\pi}\psd^2\,X_{-1}.
\label{eq:redP}
\end{equation}
Once the pressure is known, extensive variables such as the entropy and the heat capacity can be deduced. The former is derived from the entropy per unit area $s=(\partial P/\partial T)_\mu$, and the latter from the internal energy $C=(\partial U/\partial T)_\mu$. Expressed in terms of previously determined thermodynamic quantities, we have
\begin{equation}
\frac{S}{N\,\kb}=2\frac{\mcp}{\psd}-\alpha\quad\mathrm{and}\quad\frac{C}{N\,\kb}=\frac{\mcp}{\psd}-\alpha+\alpha\,X_{-1}\,X_1
\end{equation}

\end{document}